\documentclass{article}
\pdfoutput=1
\usepackage[utf8]{inputenc}
\usepackage[T1]{fontenc}
\usepackage{fixltx2e}
\usepackage{graphicx}
\usepackage{longtable}
\usepackage{float}
\usepackage{wrapfig}
\usepackage{soul}
\usepackage{textcomp}
\usepackage{marvosym}
\usepackage{latexsym}
\usepackage{amssymb}
\usepackage{hyperref}
\author{Christian Decker\\{ETH Zurich}, Switzerland\\ cdecker@tik.ee.ethz.ch \and Roger Wattenhofer\\ ETH Zurich, Switzerland\\ wattenhofer@ethz.ch}
\usepackage{todonotes}
\usepackage{cleveref}
\usepackage{listings}
\newcommand{\feb}{February }
\renewcommand{\todo}[1]{ }

\title{Bitcoin Transaction Malleability and MtGox}
\date{}
\hypersetup{
  pdfkeywords={},
  pdfsubject={},
  pdfcreator={Emacs Org-mode version 7.8.02}}

\begin{document}

\maketitle

\begin{abstract}

In Bitcoin, transaction malleability describes the fact that the
signatures that prove the ownership of bitcoins being transferred in a
transaction do not provide any integrity guarantee for the signatures
themselves. This allows an attacker to mount a malleability attack in
which it intercepts, modifies, and rebroadcasts a transaction, causing
the transaction issuer to believe that the original transaction was not
confirmed. In \feb 2014 MtGox, once the largest Bitcoin exchange, closed
and filed for bankruptcy claiming that attackers used malleability
attacks to drain its accounts. In this work we use traces of the
Bitcoin network for over a year preceding the filing to show that,
while the problem is real, there was no widespread use of malleability
attacks before the closure of MtGox.

\end{abstract}

\section{Introduction}
\label{sec-1}

In recent years Bitcoin~\cite{nakamoto2008bitcoin} has gone from a
little experiment by tech enthusiasts to a global phenomenon. The
cryptocurrency is seeing a rapid increase in adoption as well as in
value. Bitcoin is inching closer to the stated goal of creating a truly
decentralized global currency that facilitates international trade.

\todo{Bitcoin exchanges as facilitators for growth} A major
contribution of the success that Bitcoin is having today has to be
attributed to the emergence of Bitcoin exchanges. A Bitcoin exchange
is a platform that facilitates buying and selling bitcoins for
fiat money like US dollars. This enables a larger public to
come in contact with bitcoins, increasing their value as a means to
pay for goods and services. Exchanges also provide the ground truth
for the value of bitcoins by publishing their trade book and allowing
market dynamics to find a price for the traded bitcoins. Finally, much
of the media attention focuses on the rapid gain in value that these
services have enabled.

\todo{Bitcoin exchanges as single point of failure} However,
centralized exchanges are also potential points of failure, in a
system that is otherwise completely decentralized. Several high value
thefts from these services have made the headlines, never failing to
predict the impending doom of Bitcoin as a whole. Additionally a small
and mostly sentiment driven market, combined with a quick and easy way
to buy and sell bitcoins, facilitates flash crashes and rapid rallies
for no apparent reason.

\todo{MtGox History + MtGox Demise} The first, and for a long time
largest, Bitcoin exchange was MtGox. Founded in 2010 it was a first
stop for many early adopters. With the creation of other exchanges its
monopoly slowly faded, but in \feb 2014 it still accounted for
close to 70\% of all bitcoins ever traded. In \feb
2014 MtGox had to file for bankruptcy and suspend operations
following the loss of over 500 million USD worth of bitcoins owned by
its customers.

\todo{Transaction malleability} As the principal cause for the loss,
MtGox cited a problem in the Bitcoin protocol: \emph{transaction malleability}. A user could request a withdrawal from MtGox to a
Bitcoin address. The exchange would then create a corresponding
transaction and publish it to the Bitcoin network. Due to the way
MtGox tracked confirmation of these transactions it could be tricked,
exploiting transaction malleability, into believing the transaction to
have failed even though it was later confirmed by the network. MtGox
would then credit the amount back to the user's account. Effectively
the user would have doubled the withdrawn bitcoins, once from the
withdrawal and once on its account on MtGox.

\todo{In this work} In this work we investigate two fundamental
questions: Is transaction malleability being exploited? And is
the claim that it has been used to bring down MtGox plausible?
\section{Transaction Malleability}
\label{sec-2}

The Bitcoin network is a distributed network of computer nodes
controlled by a multitude of owners. They collectively implement a
replicated ledger that tracks the address balances of all users. Each
user may create an arbitrary number of addresses that can be used to
send and receive bitcoins. An address is derived from an ECDSA key pair
that is later used to prove ownership of the bitcoins associated with
that address.

The only operation allowed to modify address balances are
\emph{transactions}. A transaction is a signed data structure that on the
one hand claims some bitcoins associated with a sending address and on
the other hand reassigns them to receiving addresses. Transactions are
identified by the SHA256 hash of their serialized representation. A
transaction consists of one or more \emph{inputs} and an ordered list of
one or more \emph{outputs}. An input is used to specify which bitcoins will
be transferred, while an output specifies the address that should be
credited with the bitcoins being transferred. Formally, an output is a
tuple comprising the value that is to be transferred and a \emph{claiming condition}, expressed in a simple scripting language. An input includes
the hash of a previous transaction, an index, and a \emph{claiming script}. The hash and index form a reference that uniquely identifies
the output to be claimed and the claiming script proves that the user
creating the transaction is indeed the owner of the bitcoins being
claimed.
\subsection{Bitcoin Scripts}
\label{sec-2-1}

The scripting language is a, purposefully non-Turing complete,
stack-based language that uses single byte opcodes. The use of the
scripting language to set up both the claiming conditions and the
claiming scripts allows the creation of complex scenarios for the
transfer of bitcoins. For example, it is possible to create
multi-signature addresses that require \emph{m}-of-\emph{n} signatures
to spend the associated bitcoins for arbitration purposes. However,
the vast majority of transactions use standard scripts that set up a
claiming condition requiring the claiming script to provide a public
key matching the address and a valid signature of the current
transaction matching the public key. For this reason the standard
claiming script is generally referred to as \emph{scriptSig} (a script
encoding a signature), whereas the standard claiming condition is
referred to as \emph{scriptPubKey} (a script requiring a public key and a
signature). \Cref{fig:standard_scripts} shows the structure of the
standard claiming condition (scriptPubKey) as well as the standard
claiming script (scriptSig).

Of particular interest in this work are the
\lstinline{OP_PUSHDATA} operations which specify a number of following
bytes to be pushed as a string on the stack. Depending on the length
of the string one of several
possible flavors may be used. The simplest is a single byte with value
between $0x00$ and $0x4b$, also called \lstinline{OP_0} which simply
encodes the length of the string in itself. Additionally, three other
operations allow pushing data on the stack, namely
\lstinline{OP_PUSHDATA1}, \lstinline{OP_PUSHDATA2} and
\lstinline{OP_PUSHDATA4}, each followed by 1, 2 or 4 bytes, respectively,
encoding a little endian number of bytes to be read and pushed on the
stack.

In order to verify the validity of a transaction $t_1$ claiming an
output of a previous transaction $t_0$ the scriptSig of
$t_1$ and the scriptPubKey specified in $t_0$ are executed back
to back, without clearing the stack in between. The scriptSig of
$t_1$ pushes the signature and the public key on the stack. The
scriptPubKey of $t_0$ then duplicates the public key
(\lstinline{OP_DUP}) and replaces the first copy with its RIPEMD160
hash (\lstinline{OP_HASH160}), this 20 byte derivative of the public
key is also encoded in the address. The address from the scriptPubKey is then
pushed on the stack and the two top elements are then tested for
equality (\lstinline{OP_EQUALVERIFY}). If the hash of the public key
and the expected hash match, the script continues, otherwise execution
is aborted. Finally, the two elements remaining on the stack, i.e., the
signature and the public key, are used to verify that the signature
signs $t_1$ (\lstinline{OP_CHECKSIG}).

\begin{figure}
\begin{minipage}[h]{0.47\textwidth}
\begin{lstlisting}[frame=single, caption=\small{scriptPubKey}]
  OP_DUP
  OP_HASH160
  OP_PUSHDATA*
  <pubKeyHash>
  OP_EQUALVERIFY
  OP_CHECKSIG
\end{lstlisting}
\end{minipage}
\hfill
\begin{minipage}[h]{0.47\textwidth}
\begin{lstlisting}[frame=single, caption=\small{scriptSig}]
  OP_PUSHDATA*
  <sig>
  OP_PUSHDATA*
  <pubKey>
\end{lstlisting}
\end{minipage}
\caption{The standard claiming condition and claiming script as used
by simple transactions transferring bitcoins to an address backed by a
single public key.}
\label{fig:standard_scripts}
\end{figure}

Notice that, although the scriptSigs are attached to the inputs of the
transaction, they are not yet known at the time the signature is
created. In fact a signature may not sign any data structure
containing itself as this would create a circular dependency. For this
reason all the claiming scripts are set to a script consisting only of
a single \lstinline{OP_0} that pushes an empty string on the
stack. The user signing the transaction then iterates through the
inputs, temporarily replaces the scriptSig field with the
corresponding scriptPubKey\footnote{The use of the scriptPubKey in the signed data
as placeholder for the scriptSig is likely to avoid collisions. } from the
referenced output, and creates a signature for the resulting
serialized transaction. The signatures are then collected and inserted
at their respective positions before broadcasting the transaction to
the network.

The fact that the integrity of the scriptSig cannot be verified by
the signature is the source for transaction malleability: the claiming
script may be encoded in several different ways that do not directly
invalidate the signature itself. A simple example replaces the
\lstinline{OP_0} that pushes the public key on the stack with
\lstinline{OP_PUSHDATA2} followed by the original length. The claiming
script is changed from \lstinline{0x48<sig>41<pubKey>} to
\lstinline{0x4D4800<sig>4D4100<pubKey>}. The encoded signature is
valid in both cases but the hash identifying the transaction is
different.

Besides these changes in the way pushes are encoded, there are
numerous sources of malleability in the claiming script. A Bitcoin
Improvement Proposal (BIP) by Wuille~\cite{bip0062} identifies the
following possible ways to modify the signature and therefore exploit
malleability:

\label{sec:malleability}

\begin{enumerate}
\item ECDSA signature malleability: signatures describe points on an
    elliptic curve. Starting from a signature it is trivial to
    mathematically derive a second set of parameters encoding the same
    point on the elliptic curve;
\item Non-DER encoded ECDSA signatures: the cryptographic library used
    by the Bitcoin Core client, OpenSSL, accepts a multitude of
    formats besides the standardized DER (Distinguished Encoding
    Rules) encoding;
\item Extra data pushes: a scriptPubKey may push additional data at the
    beginning of the script. These are not consumed by the
    corresponding claiming condition and are left on the stack after
    script termination;
\item The signature and public key may result from a more complex script
    that does not directly push them on the stack, but calculates them
    on the fly, e.g., concatenating two halves of a public key that
    have been pushed individually;
\item Non-minimal encoding of push operations: as mentioned before there
    are several options to specify identical pushes of data on the
    stack;
\item Zero-padded number pushes: excessive padding of strings that are
    interpreted as numbers;
\item Data ignored by scripts: if data pushed on the stack is ignored by
    the scriptPubKey, e.g., if the scriptPubKey contains an
    \lstinline{OP_DROP}, the corresponding push in the scriptSig is
    ignored;
\item Sighash flags can be used to ignore certain parts of a script when
    signing;
\item Any user with access to the private key may generate an arbitrary
    number of valid signatures as the ECDSA signing process uses a
    random number generator to create signatures;
\end{enumerate}
\subsection{Malleability attacks}
\label{sec-2-2}

One of the problems that Bitcoin sets out to solve is the problem of
\emph{double spending}. If an output is claimed by two or more
transactions, these transactions are said to \emph{conflict}, since only
one of them may be valid. A \emph{double spending attack} is the
intentional creation of two conflicting transactions that attempt to
spend the same funds in order to defraud a third party.

\todo{Ordinary double spend attacks} Research so far has concentrated
on a classical version of the double spending attack. An attacker
would create two transactions: (1) a transaction that transfers some
of its funds once to a vendor accepting bitcoins and (2) a transaction
that transfers those same funds back to itself. The goal would then be
to convince the vendor that it received the funds, triggering a
transfer of goods or services from the vendor to the attacker, and
ensuring that the transaction returning the funds to the attacker is
later confirmed. This would defraud the vendor as the transfer to the
vendor would not be confirmed, yet the attacker received the goods or
services.

\todo{Malleability attacks}A \emph{malleability attack}, while a variant of
the double spending attack, is different from the above. The attacker
no longer is the party issuing the transaction, instead it is the
receiving party. The attacker would cause the victim to create a
transaction that transfers some funds to an address controlled by the
attacker. The attacker then waits for the transaction to be broadcast
in the network. Once the attacker has received a copy of the
transaction, the transaction is then modified using one of the above
ways to alter the signature without invalidating it. The modification
results in a different transaction identification hash. The modified
transaction is then also broadcast in the network. Either of the two
transactions may later be confirmed.

A malleability attack is said to be successful if the modified
version of the transaction is later confirmed. The mechanics of how
transactions are confirmed are complex and are out of scope for this
work. For our purposes it suffices to say that the probability of a
malleability attack to be successful depends on the distribution of
nodes in the Bitcoin network first seeing either of the transactions
(cf. \cite{bamert2013snack,decker2013information,karame2012two}). So
far the attack has not caused any damage to the victim. To be
exploitable the victim also has to rely solely on the transaction
identity hash to track and verify its account balance. Should a
malleability attack be successful the victim will only see that the
transaction it issued has not been confirmed, crediting the amount to
the attacker or attempting to send another transaction at a later
time. The attacker would have effectively doubled the bitcoins the
victim sent it.

It is worth noting that the reference client (Bitcoin Core) is not
susceptible to this attack as it tracks the unspent transaction output
set by applying all confirmed transactions to it, rather than
inferring only from transactions it issued.
\section{MtGox Incident Timeline}
\label{sec-3}

In this section we briefly describe the timeline of the incident that
eventually led to the filing for bankruptcy of MtGox. The timeline is
reconstructed from a series of press release by MtGox as well as the
official filings and legal documents following the closure.

\todo{February 7} Following several months of problems with Bitcoin
withdrawals from users, MtGox
announced~\cite{mtgox2014withdrawals} on \feb 7 that it would
suspend bitcoin withdrawals altogether. The main problem with
withdrawals was that the associated Bitcoin transactions would not be
confirmed. After this press release it was still possible to trade
bitcoins on MtGox, but it was not possible to withdraw any bitcoins
from the exchange. Specifically \cite{mtgox2014withdrawals} does not
mention transaction malleability.

\todo{Effect for users} In order to trade on MtGox, users had
transferred bitcoins and US dollars to accounts owned by MtGox. Each
user would have a virtual account that is credited with the
transferred amounts at MtGox. The withdrawal stop therefore denied
users access to their own bitcoins. While fiat currency was still
withdrawable, such a withdrawal involved a long process that would
sometimes fail altogether.

\todo{February 10} The first press release was followed by a second
press release~\cite{mtgox2014malleability} on
\feb 10, 2014. This press release claims that the problem for the
non-confirming withdrawal transactions has been identified and names
transaction malleability as the sole cause:

\begin{quote}

``Addressing Transaction Malleability: MtGox has detected unusual
activity on its Bitcoin wallets and performed investigations during
the past weeks. This confirmed the presence of transactions which need
to be examined more closely.

Non-technical Explanation: A bug in the bitcoin software makes it
possible for someone to use the Bitcoin network to alter transaction
details to make it seem like a sending of bitcoins to a bitcoin wallet
did not occur when in fact it did occur. Since the transaction appears
as if it has not proceeded correctly, the bitcoins may be
resent. MtGox is working with the Bitcoin core development team and
others to mitigate this issue.''
\end{quote}

Allegedly a user of MtGox would request a withdrawal and listen for
the resulting transaction. The transaction would then be intercepted
and replaced by a modified version that would then race with the
original transaction to be confirmed. Should the original transaction
be confirmed, the user would receive its balance only once, but not
lose any bitcoins by doing so. Should the modified transaction be
confirmed, then the user would receive the bitcoins twice: once via
the modified withdrawal transaction and a second time when MtGox
realized that the original withdrawal transaction would not confirm
and credit the users account. Implicitly in this press release MtGox
admits to using a custom client that tracks transaction validity only
via its hash, hence being vulnerable to the transaction malleability
attack.

\todo{Closure} Two more press releases followed on \feb 17 and \feb
20, both claiming that the withdrawals would resume shortly and that a
solution had been found that would eliminate the vulnerability to
malleability attacks. On \feb 23 the website of MtGox returned only
a blank page, without any further explanation, resulting in a trading
halt and the complete disappearance of MtGox. Finally on \feb 28 MtGox
announced during a press conference that it would be filing for
bankruptcy in Japan and in the
USA~\cite{mtgox2014bankruptcy,mtgox2014chapter15}.
\section{Measurements}
\label{sec-4}

Due to the nature of double spending attacks, they may only be detected
while participating in the network. As soon as one of the two
conflicting transactions is considered to be confirmed the nodes will
drop all other conflicting transactions, losing all information about
the double spending attack. Malleability attacks being a subset of
double spending attacks suffer from the same limitation.

We created specialized nodes that would trace and dump all
transactions and blocks from the Bitcoin network. These include all
double spending attacks that have been forwarded to any of the peers
our nodes connected to. Our collection of transactions started in
January 2013. As such we are unable to reproduce any attacks before
January 2013. The following observations therefore do not consider
attacks that may have happened before our collection started.

Our nodes were instructed to keep connection pools of 1,000
connections open to peers in the Bitcoin network. On average we
connected to 992 peers, which at the time of writing is approximately
20\% of the reachable nodes. According to Bamert et
al.~\cite{bamert2013snack} the probability of detecting a double
spending attack quickly converges to 1 as the number of sampled peers
increases. We therefore feel justified in assuming that the
transactions collected during the measurements faithfully reflect the
double spending attacks in the network during the same period.

Given the set of all transactions, the first task is to extract all
potential double spend attacks. In general double spending attacks can
be identified by associating a transaction with each output that it
claims. Should there be more than one transaction associated with the
same output the transactions conflict. The malleability attack being a
specialized case of the double spend attack could also be identified
by this generic procedure, however we opted for a simpler
process. Removing the signature script from a transaction results in
the signed part of the transaction, forcing all malleability attacks
to produce the same unique key. The unique key is then used to group
transactions together into \emph{conflict sets}.

\todo{General statistics} During the measurement period a total of
35,202 conflict sets were identified, each evidence of a malleability
attack. Out of these conflict sets 29,139 contained a transaction that
would later be confirmed by a block. The remaining 6,063 transactions
were either invalid because they claimed non-existing outputs, had
incorrect signatures, or they were part of a further double spending.

The \emph{conflict set value} is defined as the number of bitcoins
transferred by any one transaction in the conflict set. The outputs of
the transactions in a conflict set are identical, since any change to
them would require a new signature. In particular the value of outputs
may not be changed. Each transaction in a conflict set therefore
transfers an identical amount of bitcoins. Summing the value of all
conflict sets results in a total of 302,700 bitcoins that were
involved in malleability attacks.

\todo{Types of double spends} As mentioned in \Cref{sec:malleability},
there are a multitude of ways to use the malleability in the signature
encoding to mount a malleability attack. The most prominent type of
modification was replacing the single byte \lstinline{OP_0} with
\lstinline{OP_PUSHDATA2} which then encodes the length of the data to
push on the stack with 2 bytes. The resulting signature script would
be 4 bytes longer, because two strings are usually pushed on the
stack, but would still encode the same DER encoded signature and the
same public key, hence still be valid. A total of 28,595 out of the
29,139 confirmed attacks had this type of modifications. For the
remaining 544 conflict sets we were unable to identify the original
transactions. All transactions in these conflict sets had genuine
signatures with the correct opcodes and did not encode the same
signature. We therefore believe these transactions to be the result of
users signing raw transactions multiple times, e.g., for development
purposes.

\todo{Success rate of attacks} In order for a malleability attack to
be exploitable two conditions have to be fulfilled: (a) the modified
transaction has to be later confirmed and (b) the system issuing the
transaction must rely solely on the transaction's original hash to
track its confirmation. The first condition can be easily
reconstructed from the network trace and the Bitcoin blockchain since
only one of the transactions will be included in the
blockchain. The second condition is not detectable in our traces since
it depends on the implementation of the issuing system. In particular,
it is not possible to determine whether two payments with the same
value to the same address were intended as two separate payments or
whether an automated system issued the second one believing the first
to be invalid.

We call a malleability attack successful if it resulted in the
modified transaction to be later confirmed in a block, i.e., when
condition (a) holds. From the data derived from the attack
classification we can measure the rate of successful malleability
attacks. Out of the 28,595 malleability attacks that used an
\lstinline{OP_PUSHDATA2} instead of the default \lstinline{OP_0} only
5,670 were successful, i.e., 19.46\% of modified transactions were
later confirmed. Considering the value in malleable transactions the
success rate is comparable with 21.36\%. This reduces the total profit
of the successful attacks from 302,700 to 64,564. The strong bias
towards the original transaction is explained by the fact that the
probability of being confirmed depends on the distribution of the
transaction in the network~\cite{bamert2013snack}. During a
malleability attack the attacker listens for an incoming transaction
that match its address, modifies it and redistributes it. In the meantime
however the original transaction has been further forwarded in the
network and the modified transaction is not forwarded by nodes seeing
the original transaction. The attacker must connect to a large sample
of nodes in the network for two reasons: (a) intercept the original
transaction as soon as possible and (b) compensate the head start that
the original transaction has compared to the modified transaction.

\todo{Possible sources of double spends} So far we assumed that the
conflict sets were a direct result of a targeted attack by an attacker
against a service. There are however other causes for this kind of
conflict that should not go unmentioned. An automated system may
inadvertently create, sign a transaction and broadcast a transaction
multiple times. Due to a random parameter in the signing process the
system would produce a different signature each time, causing the
conflict that we detected. This appears to be the case with
transactions having conflict set cardinality larger than 2, that would
often not be confirmed.
\subsection{The MtGox Incident}
\label{sec-4-1}

\begin{figure}[t]
\includegraphics[width=0.5\textwidth]{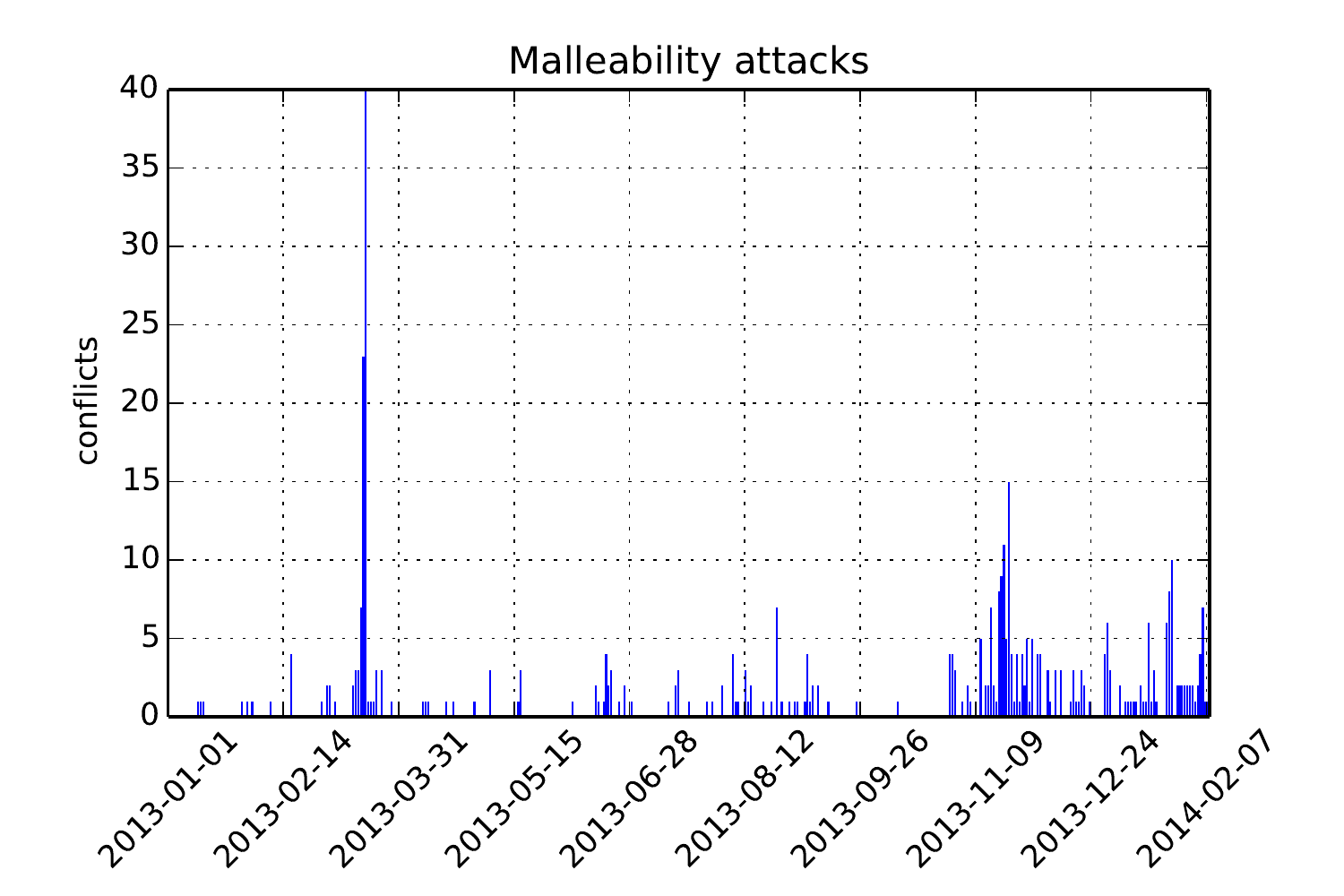}
\includegraphics[width=0.5\textwidth]{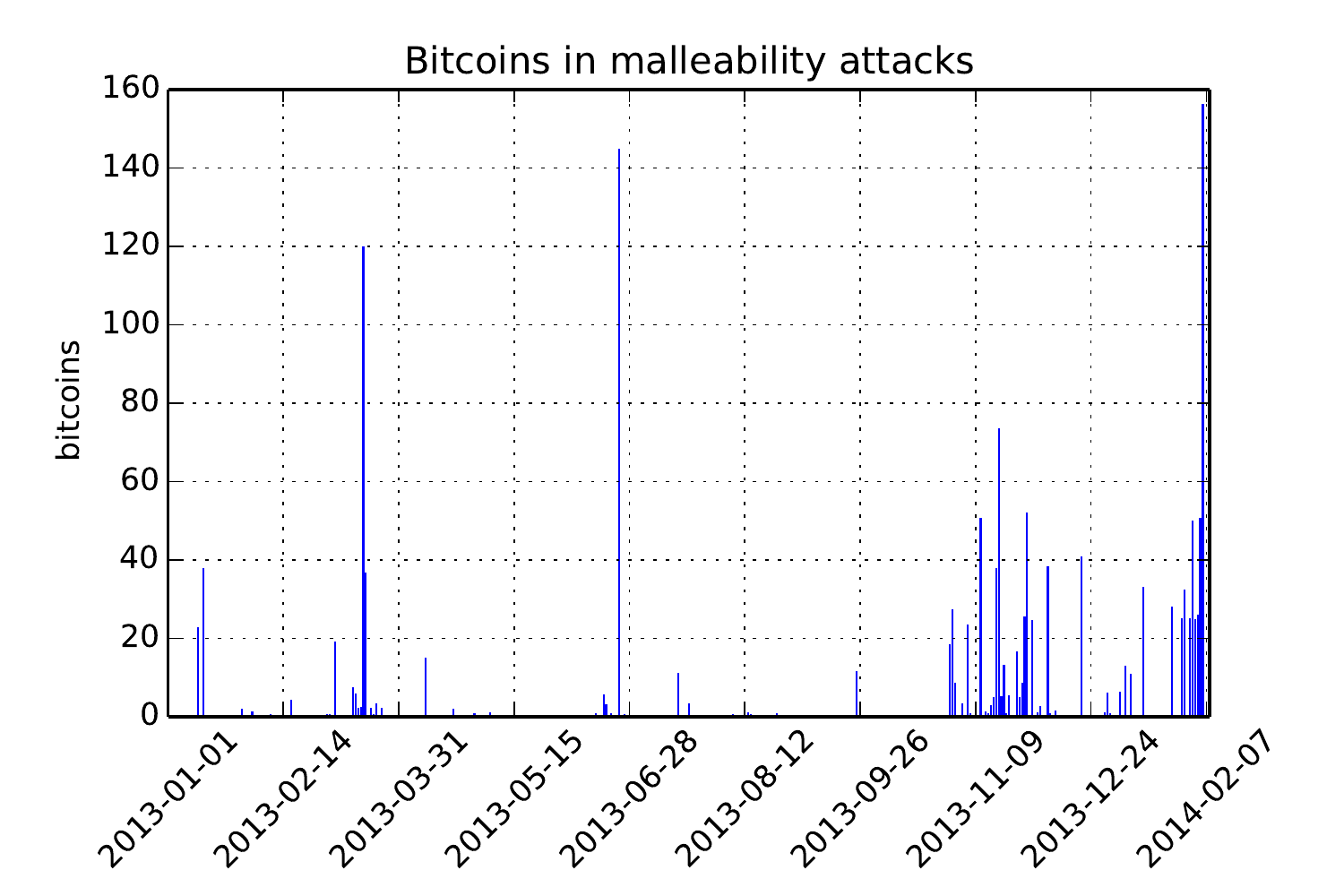}
\caption{Malleability attacks during period 1, before the
press release blaming transaction malleability as the sole cause of
losses.}
\label{fig:period1}
\end{figure}

\todo{Analyzing the MtGox breach} Returning to the specific case of
the MtGox incident of \feb 2014, that eventually lead to the closure
and the bankruptcy filing later that same month. In the press release
of \feb 10, the transaction malleability bug was explicitly named as
the root cause of the loss. The loss is later detailed as amounting to
over 850,000 bitcoins, of which 750,000 bitcoins were customer owned
bitcoins that were managed by MtGox. At the time of the first press
release bitcoins were trading at 827 US Dollars per bitcoin,\footnote{Exchange rate taken as the open value on MtGox of
\feb 7, 2014. }
resulting in a total value of lost bitcoins of 620 million US Dollars.

Assuming transaction malleability has indeed been used to defraud
MtGox, then we should be able to verify the claim by finding the
transactions used for the attack in our dataset. The above mentioned
total amount of 302,700 bitcoins involved in malleability attacks
already disproves the existence of such a large scale attack. However,
it could well be that malleability attacks contributed considerably in
the declared losses.

Reconstructing the timeline of the attacks from the announcements made
by MtGox we identify 3 time periods:

\begin{itemize}
\item Period 1 (January 2013 --- \feb 7, 2014): over a year of
   measurements until the closure of withdrawals from MtGox;
\item Period 2 (\feb 8 --- \feb 9, 2014): withdrawals are stopped
   but no details about the attack known to the public;
\item Period 3 (\feb 10 --- \feb 28): time following the
   press release blaming transaction malleability as the root cause of
   the missing bitcoins until MtGox filed for bankruptcy.
\end{itemize}

Malleability attacks in period 2 and 3 could not contribute to the
losses declared by MtGox since they happened after withdrawals have
been stopped. \Cref{fig:period1} visualizes both the number of
bitcoins involved in malleability attacks as well as the number of
attacks during period 1. During this period a total of 421 conflict
sets were identified for a total value of 1,811.58 bitcoins involved
in these attacks. In combination with the above mentioned success rate
of malleability attacks we conclude that overall malleability
attacks did not have any substantial influence in the loss of bitcoins
incurred by MtGox.

\begin{figure}
\includegraphics[width=\textwidth]{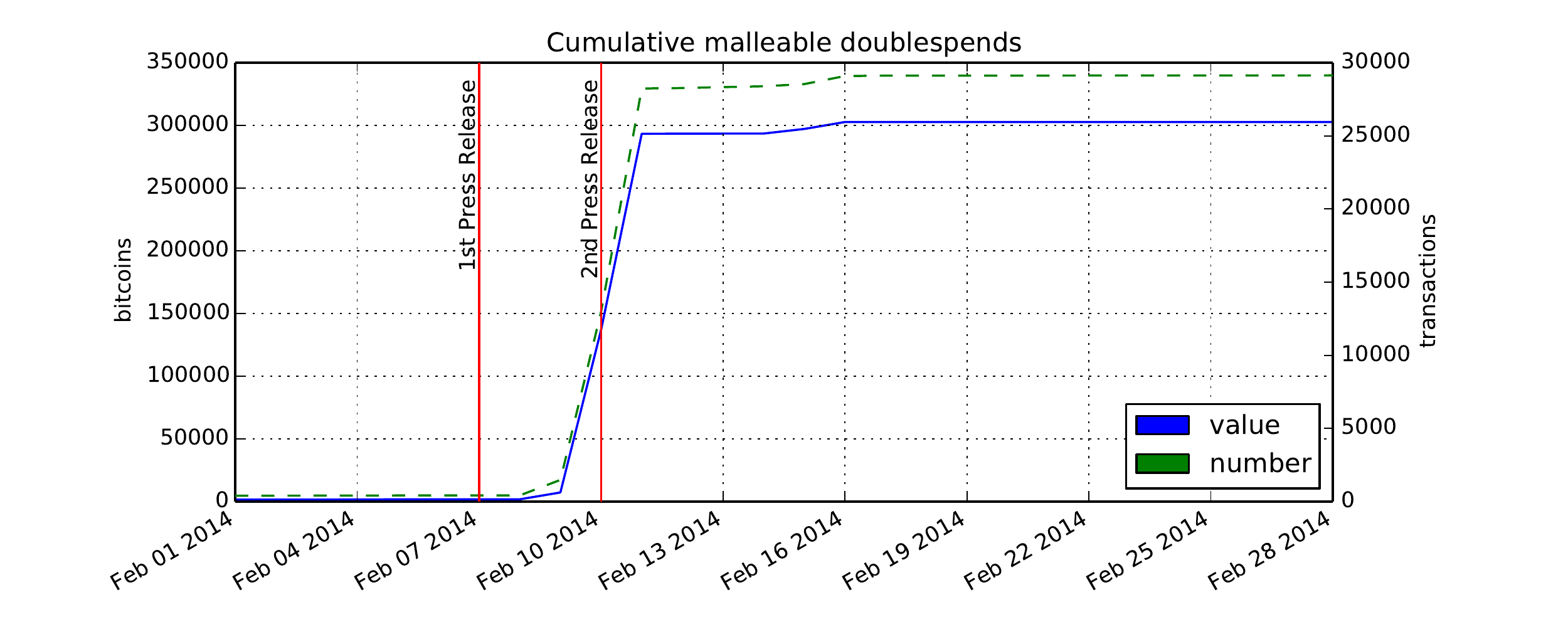}
\caption{Cumulative graph of the number and value of malleability
attacks during the time of the press releases.}
\label{fig:copycats}
\end{figure}

During period 2, we gathered 1,062 conflict sets, totalling 5,470
bitcoins. A noticeable increase of attacks at 17:00 UTC on \feb 9,
from 0.15 attacks per hour to 132 attacks per hour. While we do not
have any information about the time the second press release has been
published, the measured increase in attacks at 17:00 UTC and the date
on the press release, hints at a time between 0:00 and 2:00 JST. The
sudden increase suggests that immediately following the press release
other attackers started imitating the attack, attempting to exploit the
same weakness that had allegedly been used against MtGox.

After the second press release, in period 3, there is a sudden spike
in activity. Between \feb 10 and 11 we identified 25,752
individual attacks totalling 286,076 bitcoins, two orders of magnitude larger
than all attacks from period 1 combined. A second, smaller, wave of attacks
starts after \feb 15, with a total of 9,193 bitcoins. The attacks have
since calmed, returning to levels comparable to those observed in
period 1, before the press releases. \Cref{fig:copycats} summarizes
the situation by plotting the cumulative value and number of
malleability attacks in February 2014, i.e., from the end of
period 1 to period 3.

The strong correlation between the press releases and the ensuing
attacks attempting to exploit the same weakness is a strong indicator
that the attacks were indeed triggered by the press releases.

Assuming MtGox had disabled withdrawals like they stated in the first
press release, these attacks can not have been aimed at MtGox. The
attacks therefore where either attempts to investigate transaction
malleability or they were aimed at other businesses attempting to
imitate the purveyed attack for personal gain. The sheer amount of
bitcoins involved in malleability attacks would suggest that the
latter motive was prevalent.

It remains questionable whether other services have been informed by
MtGox in time to brace for the sudden increase in malleability
attacks. Should this not be the case then the press release may have
harmed other businesses by triggering imitators to attack them.
\section{Related Work}
\label{sec-5}

Transaction malleability has been known about since at least 2010,
when it was first documented. It has however received very little
attention so far as it was categorized as a low priority issue.

Andrychowicz et
al.~\cite{andrychowicz2013fair,andrychowicz2013deal} mention
transaction malleability as a potential problem in contracts and two
party computations based on Bitcoin transactions. These schemes can be
used for example to implement a fair coin
toss~\cite{back2014note}, auctions or decentralized voting. Their
method to eliminate transaction malleability in their protocols
resembles our construction of conflict sets, i.e., eliminating
malleable parts of the transaction in the hash calculation. However,
they limit their observations to advanced schemes for encoding
contracts and two party computations.

A related class of doublespending attacks, which we shall refer to as
classical doublespending, has received far more attention. In this
class of attacks the transaction issuer creates two transactions to
defraud the receiving party. Karame et al.~\cite{karame2012two}
first studied the problem of arising from fast transactions, i.e.,
accepting non-confirmed transactions.
Rosenfeld~\cite{rosenfeld2012analysis} showed that the success
probability of a doublespending attack can be further increased if
coupled with computational resources. Bamert et
al.~\cite{bamert2013snack} later improved the security of
accepting fast payments by observing how transactions are propagated
in the network.

To the best of our knowledge this paper is the first publication
describing transaction malleability and the resulting malleability
attack in detail. 
\section{Conclusion}
\label{sec-6}

The transaction malleability problem is real and should be considered
when implementing Bitcoin clients. 
However, while MtGox claimed to have lost 850,000 bitcoins due to malleability
attacks, we merely observed a total of 302,000 bitcoins ever being
involved in malleability attacks. Of these, only 1,811 bitcoins were in
attacks before MtGox stopped users from withdrawing bitcoins. Even more,
78.64\% of these attacks were ineffective. As such, barely 386 bitcoins could
have been stolen using malleability attacks from MtGox or from other
businesses. Even if all of these attacks were targeted against MtGox,
MtGox needs to explain the whereabouts of 849,600 bitcoins.

\bibliographystyle{plain}
\bibliography{references}

\end{document}